\documentclass[prd,twocolumn,draft]{revtex4}
\usepackage{bm}
\usepackage{hyperref}
\usepackage{latexsym}
\usepackage{mathrsfs}
\usepackage{graphicx}
\usepackage{graphics}
\usepackage{txfonts}
\usepackage{fancybox}
\usepackage{helvet}
\usepackage{fancyhdr}

\font\titlefont=cmssbx18 %at 18pt
\font\subtitlefont=cmssbx10
\font\authorfont=cmr12
\font\abstractfont=cmr10

\font\sectionfont=cmssbx10
\font\subsectionfont=cmssbx9

\font\tabletextfont=cmss10
\font\reftitlefont=cmssbx10 at 12pt
\font\reftextfont=cmr10

\begin{document}

\title{{\titlefont New Curved Spacetime Dirac Equations}\\ {\subtitlefont On the Anomalous Gyromagnetic Ratio} }
\author{\authorfont G. G. Nyambuya} 
\email{gadzirai@gmail.com, fskggn@puk.ac.za}
\affiliation{%
{\authorfont
North-West University -Potchefstroom Campus, School of Physics - Unit for Space Research, P. Bag X6001, Potchefstroom, Republic of South Africa.} \\} 

\date{\today}

\author{G. G. Nyambuya}

\begin{abstract}
{\noindent\footnotesize \textbf{Submitted:} 25 February 2008.}\\
{\footnotesize\textbf{Accepted\,\,\,:} 28 May 2008} \\
{\footnotesize\textbf{Published\,\,:} July 2008, \textsl{Foundations of Physics,} Vol. \textbf{38}, Issue 7, pages 665-677.}
\begin{center}
\linethickness{2pt}
\line(1,0){400}
\end{center}

{{\subsectionfont\noindent Abstract.} \abstractfont I propose three new curved spacetime versions of the Dirac Equation. These equations have been developed mainly to try and account in a natural way for the observed anomalous gyromagnetic ratio of Fermions. The derived equations suggest that particles including the Electron which is thought to be a point particle do have a finite spatial size which is the reason for the observed anomalous gyromagnetic ratio. A  serendipitous result of the theory, is that, two of the equation exhibits an asymmetry in their positive and negative energy solutions the first suggestion of which is clear that a solution to the problem as to why the Electron and Muon~--~despite their acute similarities~-~ exhibit an asymmetry in their mass is possible. The Mourn is often thought as an Electron in a higher energy state. Another of the consequences of three equations emanating from the asymmetric serendipity of the energy solutions of two of these equations, is that, an explanation as to why Leptons exhibit a three stage mass hierarchy is possible.}
\textbf{PACS numbers (2006):} 03.65.Pm, 11.30.j, 04.62.b, 04.62.+v, 98.80.Jk, 04.40.b, \\
\linethickness{2pt}
\line(1,0){400}
\begin{center}
\textsl{``The underlying physical laws necessary for the mathematical theory\\ of a large part of physics and the whole of chemistry are thus completely known,\\ and the difficulty is only that the exact application of these laws leads to equations\\ much too complicated to be soluble.''}
\end{center}

\begin{flushright}
-- \textbf{Paul Adrien Maurice Dirac} (1902-1984)
\end{flushright}

\end{abstract}

\maketitle

\section{\sectionfont Introduction}

%\PARstart{T}{he} 
The Dirac Equation is a relativistic quantum mechanical wave equation invented by Paul Dirac in $1928$ (Dirac 1928a, 1928b) originally designed to overcome the criticism of the Klein-Gordon Equation. The Klein-Gordon equation gave negative probabilities and this is considered to be physically meaningless. Despite this fact, this equation accounts well for Bosons, that is spin zero particles. This criticism leveled against the Klein-Gordon equation, motivated Dirac to successfully seek an equation devoid of negative probabilities.

The Dirac Equation is consistent with Quantum Mechanics (QM) and fully consistent with the Special Theory of Relativity (STR). This equation accounts in a natural way for the nature of particle spin as a relativistic phenomenon and amongst its prophetic achievements was its successful prediction of the existence of anti-particles. In its bare form, the Dirac Equation provided us with an impressive and accurate description of the Electron hence it being referred in most of the literature as the ``Dirac Equation for the Electron". It also accounts well for quarks and other spin half particles although in some of the cases, there is need for slight modifications while in others is fails - for example, one needs the Procca Equation to describe the neutron which is a spin-1/2 particle as the Electron.

The first taste of glory of the Dirac Equation was it being able to account for the gyromagnetic ratio of the electron, that is $g=2$, which can not be accounted for using non-relativistic QM. For several years after it's discovery, most physicists believed that it described the Proton and the Neutron as-well, which are both spin-1/2 particles. In simple terms, it was thought or presumed that the Dirac Equation was a universal equation for spin-1/2 particles.

However, beginning with the experiments of Stern and Frisch in $1933$, the magnetic moments of these particles were found to disagree significantly with the predictions of the Dirac Equation. The Proton was found to have a gyromagnetic ratio $g_{p}=5.58$ which is $2.79$ times larger than that predicted by the Dirac Equation. The Neutron, which is electrically neutral spin-1/2 particle was found to have a gyromagnetic ratio $g_{n}=-3.83$. 

These ``anomalous magnetic moments'' of the Neutron and Proton which are clearly not confirmatory to the Dirac Theory have been taken to be experimental indication that these partices are not fundamental particles. In the case of the Neutron, yes it is clearly not a fundamental particle since it does decay into a Proton, Electron and Neutrino, that is, $\textrm{n}\longrightarrow \textrm{p}+\textrm{e}^{-}+\nu$. If the Dirac Equation is a universal equation for fundamental fermion particles, then any fundamental fermion particle must conform to this equation.  Simple, any spin-1/2 particle that can not be described by it, must therefore not be a fundamental particle of nature.  By definition a fundamental particle is a particle known to have no sub-structure, that is, it can not be broken down into smaller particles thus will not decay into anything else.

From the Standard Model, we know that the Proton and Neutron are composed of quarks thus are not fundamental particles. The question is, is this the reason why these particle's gyromagnetic ratio is different from that predicted by the bare Dirac Equation? Prevailing wisdom suggests that anomalous gyromagnetic ratio arise because the particles under question are not fundamental particles. From the theory laid down here, the answer to this is a clear no. The reason for the deviation of the gyromagnetic ratio from that expected from the Dirac Theory is (according to this theory) because particles do have a finite size and that spacetime is curved. In this theory the anomalous gyromagnetic ratio arises from the interaction of spin with the Lorentz force in a curved spacetime for a particle of finite spatial size and mass. The derived relation for the anomalous magnetic moment and the particle size is similar to that deduced by Brodsky \& Drell (1980) and experimentally verified by Dehmelt (1989). Brodsky \& Drell (1980) proposed that fermions do have a sub-structure and this gives rise to the anomalous gyromagnetic ratio which varies as the spatial size and inverse to the mass.

\section{\sectionfont Dirac's Derivation}

Suppose we have a particle of rest mass $m_{0}$ and momentum $p$ and energy $E$, Albert Einstein, from his $1905$ special relativity paper, derived the basic equation:

\begin{equation}
E^{2}=p^{2}c^{2}+m_{0}^{2}c^{4},\label{Emc2}
\end{equation}

which later formed the basis of the Klein-Gordon Theory upon which the Dirac
Theory was founded.  This equation can be written in the matrix form:

\begin{equation}
m_{0}^{2}c^{2}=\left(
\begin{array}{c}
E/c \\
p_{x} \\
p_{y} \\
p_{z}
\end{array}\right)^{T}
\left(
\begin{array}{c c c c}
1 & 0 & 0 & 0\\
0 & -1 & 0 & 0\\
0 & 0 & -1 & 0\\
0 & 0 & 0 & -1
\end{array}\right)
\left(
\begin{array}{c}
E/c \\
p_{x} \\
p_{y} \\
p_{z}
\end{array}\right),\label{matrixform1}
\end{equation}

where the $4\times4$ matrix sandwished between the two column vectors:  

\begin{equation}
[\eta_{\mu\nu}]=\left(
\begin{array}{c c c c}
1 & 0 & 0 & 0\\
0 & -1 & 0 & 0\\
0 & 0 & -1 & 0\\
0 & 0 & 0 & -1
\end{array}\right),
\end{equation}

is the flat spacetime Minkowski metric and the superscript T in the left hand side column vector represents the transpose operation. Using the already established canonical quantisation procedures Klein and Gordon proposed the Klein-Gordon equation:

\begin{equation}
\square\Psi=\left(\frac{m_{0}c}{\hbar}\right)^{2}\Psi,\label{Klein-Gordon 1}
\end{equation}

which describes a spin-0 quantum mechanical scalar particle whose
wave-function is $\Psi$ and:

\begin{equation}
\square=\frac{1}{c^{2}}\frac{\partial^{2}}{\partial t^{2}}-\nabla^{2}.
\end{equation}

This equation allows for negative probabilities and as already stated, Dirac was not satisfied with the Klein-Gordon Theory. He noted that the Klein-Gordon equation is second order differential equation and his suspicion was that the origin of the negative probability solutions may have something to do with this very fact. He was right!

He sought an equation linear in both the time and spatial derivatives that would upon ``squaring" reproduce the Klein-Gordon equation. The equation he found was:

\begin{equation}
\left[i\hbar\gamma^{\mu}\partial_{\mu}-m_{0}c\right]\psi=0,\label{Dirac}
\end{equation}

where: 

\begin{equation}
\begin{array}{c c}
\gamma^{0}=
\left(\begin{array}{c c}
\textbf{I} & \mathbf{0}\\
\mathbf{0} & -\textbf{I} \\
\end{array}\right)
,\,\,\,\,
\gamma^{i}=
\left(\begin{array}{c c}
\textbf{0} & \mathbf{\sigma}^{i}\\
-\mathbf{\sigma}^{i} & \textbf{0} \\
\end{array}\right)
\end{array}
\end{equation}

are the  $4\times4$ Dirac gamma matrices ($\textbf{I}$ and $\textbf{0}$ are the 2$\times$2 identity and null matrices respectively) and $\psi$ is the four component Dirac wave-function. Throughout this reading, the Greek indices will be understood to mean $\mu,\nu, ... = 0,1,2,3$ and lower case English alphabet indices $i,j,k ... = 1,2,3$. 

Equation \ref{Dirac} is the original Dirac Equation. As already stated, the equation accounts very well for the Electron but is unable to account for other spin-1/2 particles without modification. Why is this so that we have to amend this beautiful equation to conform with experience? What could be the problem? Why should the equation account well for one particle and not the others? Before proceeding to derive the Dirac Equation for curved spacetime, it is instructive for latter purposes to show why the Dirac Equation is said to account very well for the Electron. 

\section{\sectionfont Dirac Gyromagnetic Ratio\label{gyro}}

We show here how the Dirac Equation discussed in the previous section accounts well for the gyromagnetic ratio of the Electron. For this reason, the Dirac Equation is said to account very well for the Electron. This discussion follows closely that of Zee (2003). In the presence of an ambient magnetic field $A_{\mu}^{ex}$, the derivatives transform as $\partial_{\mu}\longmapsto D_{\mu}=\partial_{\mu}-eA_{\mu}^{ex}$. Making this replacement results in equation \ref{Dirac} reducing to:

\begin{equation}
\left[ i\hbar\gamma^{\mu}D_{\mu}-m_{0}c \right]\psi=0\label{dirac1}.
\end{equation}

Now acting on this equation with $\left(i\hbar\gamma^{\mu}D_{\mu}+m_{0}c \right)$, we obtain $\left(\gamma^{\mu}\gamma^{\nu}D_{\mu}D_{\nu}+m^{2}_{0}c^{2}/\hbar^{2}\right)\psi=0$. We have $\gamma^{\mu}\gamma^{\nu}D_{\mu}D_{\nu}=\frac{1}{2}\left(\left\{\gamma^{\mu},\gamma^{\nu}\right\}+\left[\gamma^{\mu},\gamma^{\nu}\right]\right)D_{\mu}D_{\nu}=D_{\mu}D^{\mu}-i\sigma^{\mu\nu}D_{\mu}D_{\nu}$ and $i\sigma^{\mu\nu}D_{\mu}D_{\nu}=(i/2)\sigma^{\mu\nu}\left[D_{\mu},D_{\nu}\right]=(e/2)\sigma^{\mu\nu}F_{\mu\nu}^{ex}$ where $F_{\mu\nu}^{ex}$ is the electromagnetic field tensor of the applied external field. The above calculations reduce to:

\begin{equation}
\left(D_{\mu}D^{\mu}-\frac{e}{2}\sigma^{\mu\nu}F_{\mu\nu}^{ex}+\frac{m^{2}_{0}c^{2}}{\hbar^{2}}\right)\psi=0.
\end{equation}

Now consider a weak constant magnetic field in the z-axis such that $\vec{\textbf{A}}=(1/2)\vec{\textbf{r}}\times\vec{\textbf{B}}$ where $\vec{\textbf{B}=(0,0,B)}$ so that $F_{12}=B$. Neglecting second order terms we have

\begin{equation}
\begin{array}{c c c}
(D_{i})^{2}& = & (\partial_{i})^{2}-e(\partial_{i}A_{i}^{ex}+A_{i}^{ex}\partial_{i})+O(A_{ex,i}^{2})\\
 & =&(\partial_{i})^{2}-eB(x^{1}\partial_{2}-x^{2}\partial_{1})+O(A_{ex,i}^{2})\\
&=&\vec{\nabla}^{2}-e\vec{\textbf{B}}\cdot\vec{\textbf{L}}+O(A_{ex,i}^{2})
\end{array},
\end{equation}

where $\vec{\textbf{L}}=\vec{\textbf{r}}\times\vec{\textbf{p}}$ is the orbital angular momentum operator which means that the orbital angular momentum generates orbital magnetic moment that interacts with the magnetic field. Now, if we write the Dirac four component wave-function as $\psi=\left(\begin{array}{c}
\Phi\\
\chi
\end{array}
\right)$, we find that in the non-relativistic limit the component $\chi$ dominates. Thus, $e\sigma^{\mu\nu}F_{\mu\nu}/2$ acting on $\Phi$ is effectively equal to $\frac{e}{2}\sigma^{3}(F_{12}-F_{21})=2e\vec{\textbf{B}}\cdot\vec{\textbf{S}}$ since $\vec{\textbf{S}}=(\vec{\boldsymbol{\sigma}}/2)$. Now writing $\Phi=e^{-im_{0}t}\Psi$ where $\Psi$ oscillates much more slowly than $e^{im_{0}t}$ so that $(\partial_{0}^{2}+m^{2}_{0}c^{2}/\hbar^{2})e^{-im_{0}c^{2}t/\hbar}\Psi\simeq e^{-im_{0}c^{2}t/\hbar}\left[-(2im_{0}c/\hbar)\partial_{0}\Psi\right]$. Putting all the bits and pieces together, we have:

\begin{equation}
\left[\frac{\hbar^{2}}{2m_{0}}\vec{\nabla}^{2}+\mu_{B}\vec{\textbf{B}}\cdot(\vec{\textbf{L}}+2\vec{\textbf{S}})\right]\Psi=-i\hbar\frac{\partial \Psi}{\partial t},\label{gyro_eqn}
\end{equation}

and this equation above and below embodies the historic fit of the Dirac Equation in that it automatically tells us that the gyromagnetic ratio of the Electron is $2$. However as already explained, precise measurements put this value slightly above $2$ and this desperacy in observations and theory caused the theorist to go back to the drawing board to seek harmony with observations. With the emergence of QED, this desperacy was solved considering particle-particle interactions through the so-called Feynman diagrams/method. This approach has yielded the best ever agreement for any theory ever conceived by the human mind. The agreement between theory and observation is so impressive that QED has be dubbed ``the best theory we have''.

If QED already explains the gyromagnetic ratio, why then seek another theory that tries to explain this same phenomena? The problem is that QED assumes that the Electron and also the Muon whose gyromagnetic ratio are very close to the Dirac values of 2 are point particles. It is this very fact that this paper seeks to put back on the drawing board for a deeper introspection. Is and or are the Electron and the Muon point particles? The other is for philosophical reasons that I need not go into the details but maybe in-passing mention that from a higher level of beauty, I expect that the Dirac Equation should account naturally for all fermions without much drastic modifications - if any at all, in much the same way as it was once conceived as a universal equation for spin-1/2 particles. The other is that the methods such as renormalization that are used in QED to arriving at these all-time accurate values feel or appear rather un-natural. Yes, they yield to the right numbers when compared with reality but is such a program really part and parcel of the natural laws? This is a very difficult question to answer from an experimental point of view simple because of its philosophical nature.

\section{\sectionfont Vierbein Curved Spacetime Dirac Equation}

The Dirac Equation can be written in curved spacetime using vierbein fields as:

\begin{equation}
i\gamma^a e_a^\mu D_\mu \psi - \left(\frac{m_{0}c}{\hbar}\right) \psi = 0,\label{vdirac}
\end{equation}

and this equation is generally accepted as the Dirac Equation in curved spacetime (see e.g Lawrie 1990), where $e_a^\mu$  is the vierbein field and $D_{\mu}$ is the covariant derivative for fermion fields, defined as follows
$D_\mu = \partial_\mu - \frac{i}{4} \eta_{ac} \omega^c_{b\mu} \sigma^{ab}$ where $\eta_{ac}$ is the Lorentzian metric, $\sigma^{ab}$ is the commutator of Dirac matrices: 

\begin{equation}
\sigma^{ab}=\frac{i}{2} \left[\gamma^{a},\gamma^{b}\right]
\end{equation}

and the square brackets represent the usual commutator brackets and  $\omega^c_{b\mu}$ is the spin connection and is given by:

\begin{equation}
\omega^c_{b\mu} = e^c_\nu \partial_\mu e^\nu_b + e^c_\nu e^\sigma_b \Gamma^\nu_{\sigma\mu},
\end{equation}

and $\Gamma^\nu_{\sigma\mu}$ is the Christoffel symbol and the Latin indices denote the ``Lorentzian'' indices while the Greek indices denote here the ``Riemannian'' indices. Vierbein fields describe a local frame that enables one to define the Dirac matrices at any given point in spacetime. 

Naturally, the question arises; If equation \ref{vdirac} is a legitimate and accepted Dirac Equation for curved spacetime -- \textbf{Why seek another equation for the same curved spacetime?} We seek another equation for the same spacetime using a mathematically legitimate approach because this approach, unlike the one in the present section, leads to equations that not only explain the anomalous gyromagnetic ratio, but equations that exhibit asymmetric energy solutions and these energy solutions place us in a position to explain ``mysteries'' such as the existence of the generation of Leptons -- hence the motivation and need. Additionally, having two legitimate equations, explaining the same thing from two different approaches, means, if one measures the predictions of these equations against reality, only one of them will give predictions that corresponds to experience, hence a way to test which approach actually conforms to nature is presented. 

\section{\sectionfont New Curved Spacetime Dirac Equations}

Let us begin by looking at the Einstein equation \ref{Emc2}. We know that its equivalent in curved spacetime is given by: 

\begin{equation}
g_{\mu\nu}p^{\mu}p^{\nu}=m_{0}^{2}c^{2},\label{curved}
\end{equation}

where $p^{\mu}=(E,\textbf{p})$ is the usual four momentum and $g_{\mu\nu}$ is the metric of spacetime. Writing this in matrix form as equation \ref{matrixform1}, we have:

\begin{equation}
m_{0}^{2}c^{2}=\left(
\begin{array}{c}
E/c \\
p_{x} \\
p_{y} \\
p_{z}
\end{array}\right)^{T}
\left(
\begin{array}{c c c c}
g_{00} & g_{01} & g_{02} & g_{03}\\
g_{10} & g_{11} & g_{12} & g_{13}\\
g_{20} & g_{21} & g_{22} & g_{23}\\
g_{30} & g_{21} & g_{32} & g_{33}
\end{array}\right)
\left(
\begin{array}{c}
E/c \\
p_{x} \\
p_{y} \\
p_{z}
\end{array}\right),\label{meqn1}
\end{equation}

and we see here that there are off-diagonal terms in this expression hence thus in curved spacetime, off-diagonal ought to emerge and thus we seek an equation that is first order in its partial derivatives which upon squaring (in the Dirac sense of squaring) brings in the off diagonal terms.  In the following subsections, we will derive three such equations.

\subsection{\subsectionfont Equation I}

As a first step, we will set the off-diagonals terms to zero while introducing some curvature to equation \ref{meqn1}, that is:

\begin{equation}
m_{0}^{2}c^{2}=\left(
\begin{array}{c}
E/c \\
p_{x} \\
p_{y} \\
p_{z}
\end{array}\right)^{T}
\left(
\begin{array}{c c c c}
A_{0}A_{0} & 0 & 0 & 0\\
0 & -A_{1}A_{1} & 0 & 0\\
0 & 0 & -A_{2}A_{2} & 0\\
0 & 0 & 0 & -A_{3}A_{3}
\end{array}\right)
\left(
\begin{array}{c}
E/c \\
p_{x} \\
p_{y} \\
p_{z}
\end{array}\right),\label{meqn2}
\end{equation}

where $A_{\mu}=A_{\mu}(\hat{\textbf{r}},t)$ is well behaved vector function and the metric tensor is here given by:

\begin{equation}
g_{\mu\nu}=\left(
\begin{array}{c c c c}
A_{0}A_{0} & 0 & 0 & 0\\
0 & -A_{1}A_{1} & 0 & 0\\
0 & 0 & -A_{2}A_{2} & 0\\
0 & 0 & 0 & -A_{3}A_{3}
\end{array}\right).
\end{equation}

This function will not be defined in this reading but it will suffice to note that the metric can legally be written in this form. Now, it is a straight forward mathematical exercise that if we multiplied the left hand-side of the original Dirac Equation given in equation \ref{Dirac} by $A^{\mu}$, that is:

\begin{equation}
\left[i\hbar A^{\mu}\gamma^{\mu}\partial_{\mu}-m_{0}c\right]\psi=0,\label{cdirac1}
\end{equation}

we will recover equation \ref{meqn2} after multiplying equation \ref{cdirac1} from the left by $i\hbar A^{\mu}\gamma^{\mu}\partial_{\mu}+m_{0}c$ on the condition that $\partial_{\mu}A^{\mu}=0$. This equation is Lorentz invariant.

{\subsectionfont \underline{Lorentz Invariance}:} To prove Lorentz covariance two conditions must be satisfied:

1. Given any two inertial observers $\textrm{O}$ and $\textrm{O}^\prime$ anywhere in spacetime, if in the frame $\textrm{O}$ we have $[i\hbar A^{\mu}\gamma^{\mu}\partial_{\mu}-m_{0}c]\psi(x)=0$, then $[i\hbar A^{\prime\mu}\gamma^{\prime\mu}\partial_{\mu}^\prime-m_{0}c]\psi^\prime(x^\prime)=0$ is the equation describing the same state but in the frame $\textrm{O}^\prime$. 

2. Given that $\psi(x)$ is the wavefunction as measured by observer $\textrm{O}$, there must be a prescription for observer $\textrm{O}^\prime$ to compute $\psi^\prime(x^\prime)$ from $\psi(x)$ and this describes to $\textrm{O}^\prime$ the same physical state as that measured by $\textrm{O}$.

Now, since the Lorentz transformation are linear, it is to be required or expected of the transformations between $\psi(x)$ and $\psi^\prime(x^\prime)$  to be linear too, that is:

\begin{equation} 
\psi^\prime(x^\prime) = \psi^\prime(\Lambda x) = S(\Lambda) \psi(x) = S(\Lambda) \psi(\Lambda^{-1}x^\prime)\label{inverse1}
\end{equation} 

where $S(\Lambda)$ is a $4\times 4$ matrix which depends only on the relative velocities of $\textrm{O}$ and $\textrm{O}^\prime$ and $\Lambda$ is the Lorentz transformation matrix. $S(\Lambda)$ has an inverse if $\textrm{O}\rightarrow \textrm{O}^\prime$ and also $\textrm{O}^\prime\rightarrow \textrm{O}$. The inverse is:

\begin{equation} 
\psi(x) = S^{-1}(\Lambda)\psi^\prime(x^\prime) = S^{-1}(\Lambda)\psi^\prime(\Lambda x) \label{inverse2}
\end{equation} 	

or we could write:

\begin{equation}
\psi(x)=S(\Lambda^{-1})\psi^\prime(\Lambda x)\Longrightarrow S(\Lambda^{-1}) = S^{-1}(\Lambda)
\end{equation}

We can now write $[i\hbar A^{\mu}\gamma^{\mu}\partial_{\mu}-m_{0}c]\psi(x)=0$ 
as  $[i\hbar A^{\mu}\gamma^{\mu}\partial_{\mu}-m_{0}c]S^{-1}(\Lambda)\psi^\prime(x^\prime)=0$ and multiplying this from the left by $S(\Lambda)$ we have $S(\Lambda)[i\hbar A^{\mu}\gamma^{\mu}\partial_{\mu}-m_{0}c] S^{-1}(\Lambda)\psi^\prime(x^{\prime})=0$
and hence:

\begin{equation}
\left(i\hbar S(\Lambda)\gamma^{\mu} S^{-1}(\Lambda)A^{\mu}\partial_\mu - m_{0}c\right)\psi^\prime(x^\prime)=0.
\end{equation}

Now, since $A^{\mu}$ is a vector, it is clear that $A^{\mu}\partial_\mu$ is a scalar, that is, $A^{\mu}\partial_{\mu}=A^{\prime\mu}\partial^{\prime}_{\mu}$, therefore we will have:

\begin{equation}
\left(i\hbar S(\Lambda)\gamma^{\mu} S^{-1}(\Lambda)A^{\prime\mu}\partial^{\prime}_{\mu} - m_{0}c\right)\psi^\prime(x^\prime)=0.
\end{equation}

Therefore, for equation \ref{cdirac1} to be Lorentz invariant, the requirement is that:

\begin{equation} 
\gamma^{\prime\mu}=S(\Lambda)\gamma^{\mu} S^{-1}(\Lambda), 
\end{equation}

hence thus we have shown that equation \ref{cdirac1} is Lorentz invariant. 

{\subsectionfont \underline{Gyromagnetic Ratio}:} Following the same procedure as in section \ref{gyro}, we find that the anomalous gyromagnetic ratio ($\Delta a^{f}$) emerging from equation \ref{cdirac1} is given by:

\begin{equation}
\Delta a^{f}=\left(\frac{g-2}{2}\right)=A_{1}A_{2}-1=\sqrt{g_{11}g_{g_{22}}}-1,\label{fgratio}
\end{equation}

and this deviates from the Dirac value of $\Delta a^{f}=0$ because for a curved spacetime, $g_{11}g_{22}$ will not necessarily be equal to unity as is the case for a flat spacetime applicable to the bare Dirac Equation.

{\subsectionfont \underline{Radius of a Schwarzschild-Electron}:} Assuming that equation \ref{cdirac1} describes an Electron and that the spacetime of this Electron is described by the Schwarzschild metric:

\begin{equation}
ds^{2}=\left({1-\frac{2R_{S}}{r}}\right)c^{2}dt^{2}-\left[\left({1-\frac{2R_{S}}{r}}\right)^{-1}dr^{2}+\left(\frac{r}{R_{p}}\right)^{2}d\Omega^{2}\right],
\end{equation}

where $d\Omega^{2}=R^{2}_{p}\left(d\theta^{2}+\sin^{2}\theta d\varphi^{2}\right)$, $R_{p}$ is the particle radius and $R_{S}=Gm_{p}/c^{2}$ is Schwarzschild radius where $G$ is Newton's universal constant of gravitation and $m_{p}$ the mass of the particle, we find to first order approximation (where $\sqrt{g_{11}}\sim 1+R_{S}/r$) that:

\begin{equation}
\sqrt{g_{11}g_{22}}=\frac{r}{R_{p}}+\frac{R_{S}}{R_{p}},
\end{equation}

and evaluating this at the surface of the particle, that is, at $r=R_{p}$ and then substituting this into equation \ref{fgratio}, we are lead to: 

\begin{equation}
\Delta a^{f}=\frac{R_{S}}{R_{p}}.
\end{equation}

A similar relation to the above, namely $|g-2|=R_{p}/\lambda_{c}$, has been proposed by Brodsky $\&$ Drell (1980) to explain the origin of this ratio. These authors -- as is in the present -- take to mean that the presence of a non-zero anomalous gyromagnetic moment is due to the fact that these particles have a finite size and are not point-like (without length, height and width) as is usually assumed in QED.

Now, given the current Electron anomalous gyromagnetic ratio of $0.0011596521811\pm 0.00000000000075$ (see any good book on Quantum Electrodynamics), this could mean the radius of an Electron described by the Schwarzschild metric is about $862$ times the Electron Schwarzschild radius, that is $5.84\times10^{-55}\rm{m}$. Practically speaking, this radius is small enough to consider the Electron to be a point-particle but this does not make it a point-particle. It should be said, that, the Schwarzschild metric describes a non-rotating and non-spinning particle. The Electron certainly possess spin and rotation thus a Schwarzschild metric description of it, is an approximation and for a more accurate description, one will need a metric tensor that incooperates both spin and rotation.

\subsection{\subsectionfont Equation II}

To obtain an equation for a generally curved spacetime, we note that the metric can be written in the form:

\begin{equation}
[g_{\mu\nu}]=\left(
\begin{array}{r r r r}
A_{0}A_{0} & A_{0}A_{1} & A_{0}A_{2} & A_{0}A_{3}\\
A_{1}A_{0} & -A_{1}A_{1} & A_{1}A_{2} & A_{1}A_{3}\\
A_{2}A_{0} & A_{2}A_{1} & -A_{2}A_{2} & A_{2}A_{3}\\
A_{3}A_{0} & A_{3}A_{1} & A_{3}A_{2} & -A_{3}A_{3}
\end{array}\right)\label{pos-m}
\end{equation}

and from this, the energy equation will be given by: 

\begin{equation}
m_{0}^{2}c^{2}=\left(
\begin{array}{c}
E/c \\
p_{x} \\
p_{y} \\
p_{z}
\end{array}\right)^{T}
\left(
\begin{array}{r r r r}
A_{0}A_{0} & A_{0}A_{1} & A_{0}A_{2} & A_{0}A_{3}\\
A_{1}A_{0} & -A_{1}A_{1} & A_{1}A_{2} & A_{1}A_{3}\\
A_{2}A_{0} & A_{2}A_{1} & -A_{2}A_{2} & A_{2}A_{3}\\
A_{3}A_{0} & A_{3}A_{1} & A_{3}A_{2} & -A_{3}A_{3}
\end{array}\right)
\left(
\begin{array}{c}
E/c \\
p_{x} \\
p_{y} \\
p_{z}
\end{array}\right).\label{energy2}
\end{equation}

Now if we take the equation:

\begin{equation}
\left[i\hbar A^{\mu}\bar{\gamma}^{\mu}\partial_{\mu} -m_{0}c\right]\psi=0.\label{cdirac2}
\end{equation}

where:

\begin{equation}
\begin{array}{c c}
\bar{\gamma}^{0}=
\left(\begin{array}{c c}
\textbf{I} & \mathbf{0}\\
\mathbf{0} & -\textbf{I} \\
\end{array}\right)
,\,\,\,\,
\bar{\gamma}^{i}=
\frac{1}{2}\left(\begin{array}{c c}
2\textbf{I} & i\sqrt{2}\mathbf{\sigma}^{i}\\
-i\sqrt{2}\mathbf{\sigma}^{i} & -2\textbf{I} \\
\end{array}\right),
\end{array}\label{gamma-bar-m}
\end{equation}

it is not a difficult exercise to show that multiplication of this equation from the left handside by the operator $\left[i\hbar A^{\mu}\bar{\gamma}^{\dagger\mu}\partial_{\mu} +m_{0}c\right]$ leads us to the energy equation \ref{energy2} provided $\partial_{\mu}A^{\mu}=0$.

\subsection{\subsectionfont Equation III}

Again, we obtain another equation for a generally curved spacetime, by noting that the metric can be written in another form, namely:

\begin{equation}
[g_{\mu\nu}]=\left(
\begin{array}{r r r r}
A_{0}A_{0} & -A_{0}A_{1} & -A_{0}A_{2} & -A_{0}A_{3}\\
-A_{1}A_{0} & -A_{1}A_{1} & A_{1}A_{2} & A_{1}A_{3}\\
-A_{2}A_{0} & A_{2}A_{1} & -A_{2}A_{2} & A_{2}A_{3}\\
-A_{3}A_{0} & A_{3}A_{1} & A_{3}A_{2} & -A_{3}A_{3},
\end{array}\right)
\end{equation}

which is just the metric given in equation \ref{pos-m} but having undergone the transformation $A_{0}\longmapsto -A_{0}$ or $A_{k}\longmapsto -A_{k}$. As before, the energy equation for this metric will be given by: 

\begin{equation}
m_{0}^{2}c^{2}=\left(
\begin{array}{c}
E/c \\
p_{x} \\
p_{y} \\
p_{z}
\end{array}\right)^{T}
\left(
\begin{array}{r r r r}
A_{0}A_{0} & -A_{0}A_{1} & -A_{0}A_{2} & -A_{0}A_{3}\\
-A_{1}A_{0} & -A_{1}A_{1} & A_{1}A_{2} & A_{1}A_{3}\\
-A_{2}A_{0} & A_{2}A_{1} & -A_{2}A_{2} & A_{2}A_{3}\\
-A_{3}A_{0} & A_{3}A_{1} & A_{3}A_{2} & -A_{3}A_{3},
\end{array}\right)
\left(
\begin{array}{c}
E/c \\
p_{x} \\
p_{y} \\
p_{z}
\end{array}\right).\label{energy3}
\end{equation}

Now if we take the equation:

\begin{equation}
\left[i\hbar A^{\mu}\hat{\gamma}^{\mu}\partial_{\mu} -m_{0}c\right]\psi=0.\label{cdirac3}
\end{equation}

where:

\begin{equation}
\begin{array}{c c}
\hat{\gamma}^{0}=
\left(\begin{array}{c c}
\textbf{I} & \mathbf{0}\\
\mathbf{0} & -\textbf{I} \\
\end{array}\right)
,\,\,\,\,
\hat{\gamma}^{i}=
-\frac{1}{2}\left(\begin{array}{c c}
2\textbf{I} & i\sqrt{2}\mathbf{\sigma}^{i}\\
-i\sqrt{2}\mathbf{\sigma}^{i} & -2\textbf{I} \\
\end{array}\right),
\end{array}\label{gamma-hat-m}
\end{equation}

or 

\begin{equation}
\begin{array}{c c}
\hat{\gamma}^{0}=
-\left(\begin{array}{c c}
\textbf{I} & \mathbf{0}\\
\mathbf{0} & -\textbf{I} \\
\end{array}\right)
,\,\,\,\,
\hat{\gamma}^{i}=
\frac{1}{2}\left(\begin{array}{c c}
2\textbf{I} & i\sqrt{2}\mathbf{\sigma}^{i}\\
-i\sqrt{2}\mathbf{\sigma}^{i} & -2\textbf{I} \\
\end{array}\right).
\end{array}\label{gamma-hat-m1}
\end{equation}

Once again, it is not a difficult exercise to show that multiplication of this equation from the left handside by the operator $\left[i\hbar A^{\mu}\hat{\gamma}^{\dagger\mu}\partial_{\mu} + m_{0}c\right]$ leads us to the energy equation \ref{energy3} provided $\partial_{\mu}A^{\mu}=0$. Just as equation \ref{cdirac1}, equation \ref{cdirac2} and equation \ref{cdirac3} are Lorentz invariant and to show this, one needs to go through the same steps as those taken to show the Lorentz invariance of equation \ref{cdirac1}.

\section{\sectionfont Anomalous Gyromagnetic Ratio} 

Following the same procedure as in section \ref{gyro}, we expose the particle to an ambient magnetic field. We make the same simplification as in section \ref{gyro}. First, we multiply  equation \ref{cdirac3} for a particle inside an ambient magnetic field, that is, $\left[i\hbar A^{\mu}\bar{\gamma}^{\mu}D_{\mu}- m_{0}c\right]\psi=0$ by the operator $\left[i\hbar A^{\mu}\bar{\gamma}^{\mu\dagger}D_{\mu}+ m_{0}c\right]$  we obtain $\left(\bar{\gamma}^{\mu\dagger}\bar{\gamma}^{\nu}A^{\mu}A^{\nu}D_{\mu}D_{\nu}+m^{2}_{0}c^{2}/\hbar^{2}\right)\psi=0$. Now, $\bar{\gamma}^{\mu\dagger}\bar{\gamma}^{\nu}A^{\mu}A^{\nu}D_{\mu}D_{\nu}=\frac{1}{2}\left(\left\{\bar{\gamma}^{\mu\dagger},\bar{\gamma}^{\nu}\right\}+\left[\bar{\gamma}^{\mu\dagger},\bar{\gamma}^{\nu}\right]\right)A^{\mu}A^{\nu}D_{\mu}D_{\nu}=\bar{\eta}^{\mu\nu}A^{\mu}A^{\nu}D_{\mu}D_{\mu}-i\bar{\sigma}^{\mu\nu}A^{\mu}A^{\nu}D_{\mu}D_{\nu}$ and $i\bar{\sigma}^{\mu\nu}A^{\mu}A^{\nu}D_{\mu}D_{\nu}=(i/2)\bar{\sigma}^{\mu\nu}A^{\mu}A^{\nu}\left[D_{\mu},D_{\nu}\right]=(e/2)\bar{\sigma}^{\mu\nu}A^{\mu}A^{\nu}F_{\mu\nu}^{ex}$ where $F_{\mu\nu}^{ex}$ is the electromagnetic field tensor of the applied external field,

\begin{equation}
[\bar{\sigma}_{\mu\nu}]=\frac{1}{\sqrt{2}}\left(
\begin{array}{c c c c}
0 & \gamma^{0}\gamma^{1} & \gamma^{0}\gamma^{2} & \gamma^{0}\gamma^{3}\\
-\gamma^{0}\gamma^{1} & 0 & \gamma^{3} & \gamma^{2}\\
-\gamma^{0}\gamma^{2} & -\gamma^{3} & 0 & \gamma^{1}\\
-\gamma^{0}\gamma^{3} & -\gamma^{2} & -\gamma^{1} & 0\\
\end{array}
\right).
\end{equation}

and:

\begin{equation}
[\bar{\eta}_{\mu\nu}]=\left(
\begin{array}{rrrr}
1 & 1 & 1 & 1\\
1 & -1 & 1 & 1\\
1 & 1 & -1 & 1\\
1 & 1 & 1 & -1\\
\end{array}
\right).
\end{equation}

The above calculations reduces to:

\begin{widetext}

\begin{equation}
\left(\mathcal{D}_{\mu}\mathcal{D}^{\nu}-\frac{1}{2}\bar{\sigma}^{\mu\nu}A^{\mu}A^{\nu}\left(i[\partial_{\mu},\partial_{\nu}]+[A_{\mu}^{ex},\partial_{\nu}]+F_{\mu\nu}^{ex}\right)+\frac{m_{0}^{2}c^{2}}{\hbar^{2}}\right)\psi=0,
\end{equation}

\end{widetext}

where $\mathcal{D}_{\mu}=A^{\mu}D_{\mu}$ and $\mathcal{D}^{\mu}=\bar{\eta}^{\mu\nu}\mathcal{D}_{\nu}$. Applying the same-kind of operations to equation \ref{cdirac2}, one arrives at the same equation thus the gyromagnetic ratio to be derived from this applies to both equation \ref{cdirac2} and \ref{cdirac3}. Now, just as in section \ref{gyro} we proceed to simplify by making the same approximations, and in so doing, a new term incooperating the spin emerges from the term $e\bar{\sigma}^{01}A^{0}A^{1}F_{01}/2+e\bar{\sigma}^{10}A^{1}A^{0}F_{10}/2+e\bar{\sigma}^{02}A^{0}A^{2}F_{02}/2+e\bar{\sigma}^{20}A^{2}A^{0}F_{20}/2$ and assuming that spacetime is not severely curved ($A_{\mu}\sim1)$,  then this new term is given by: 

\begin{equation}
\frac{e\vec{\textbf{S}}\cdot\vec{\textbf{v}}\times\vec{\textbf{B}}}{c\sqrt{2}}=\frac{ev\sin\theta}{c\sqrt{2}}\vec{\textbf{B}}\cdot\vec{\textbf{S}}
\end{equation}

where $\theta$ is the angle between $\vec{\textbf{v}}$ and $\vec{\textbf{B}}$. Here to interpret $\vec{\textbf{v}}$, we envisage an Electron as a point particle orbiting a central point (while spinning on its own axis like the planets round the sun) and the size of the circle inscribed by the orbiting Electron is then the radius of the Electron! If we assume that spacetime is not severely curved, then as before, the meaning of which is that $A_{\mu}\sim1$ hence $\mathcal{D}_{\mu}\mathcal{D}^{\mu}\sim\nabla^{2}$, then all this reduces to:

\begin{equation}
\frac{\hbar^{2}}{2m_{0}}\nabla^{2}\Psi+\mu_{B}\vec{\textbf{B}}\cdot\left(\vec{\textbf{L}}+\left[\sqrt{2}+\frac{\lambda_{c}\sin\theta}{\sqrt{2}R_{p}}\right]\vec{\textbf{S}}\right)\Psi=-i\hbar\frac{\partial \Psi}{\partial t},
\end{equation}

which tells us that for every unit of spin angular momentum the spin will interact with  $\left(\sqrt{2}+\lambda_{c}\sin\theta/\sqrt{2}R_{p}\right)$ times with the magnetic field hence thus the anomalous gyromagnetic ratio for the particle in question is given by:

\begin{equation}
\Delta a^{c}=\left(\frac{g-2}{2}\right)=\frac{1}{\sqrt{2}}+\frac{\lambda_{c}\sin\theta}{2\sqrt{2}R_{p}}-1.
\end{equation}

We should note that $\lambda_{c}=\hbar/mv$ is the Compton wavelength of the particle in question and the symbols have their usual meanings. 

\section{\sectionfont Energy Solutions}

Written in a compact notation, the energy equations for the particles described by equations \ref{cdirac1}, \ref{cdirac2} and \ref{cdirac3} are:

\begin{equation}
(A^{0})^{2}E^{2}-\left(2\lambda A^{0}A^{k} p_{k}c\right)E-A^{j}A^{k}p_{j}p_{k}c^{2} = m_{0}^{2}c^{4},\label{energy1}
\end{equation}

where $\lambda=\pm1,0$ and the case $\lambda=0$ is the case for equation \ref{cdirac1}, $\lambda=+1$ is the case for a for equation \ref{cdirac2} and likewise $\lambda=-1$ is the case for equation \ref{cdirac3}. Setting $m^{*}_{0}=m_{0}/A^{0}$, $\mathcal{P}=A^{k}p_{k}/A^{0}$  and realizing that $A^{j}A^{k}p_{j}p_{k}= (A^{k}p_{k})^{2}=(A^{0})^{2}\mathcal{P}^{2}$, the solution to this energy equation (that is, equation \ref{energy1}) with respect to $E$ is given by:

\begin{equation}
E=\lambda \mathcal{P}c\pm \sqrt{\left(1+\left|\lambda\right|\right)\mathcal{P}^{2}c^{2}+\left(m^{*}_{0}c^{2}\right)^{2}}.
\end{equation}

From this, its clear that we will have three negative energy particles and three positive energy particles. 

Adopting Dirac's interpretation of the vacuum, that negative energy states are filled hence are not observable, it would mean there should according to equations \ref{cdirac1}, \ref{cdirac2} and \ref{cdirac3} exist three particles whose masses are non-identical.  We know that Lepton's come in three generations and each generation is divided into two Leptons and the two Leptons may be divided into one with electric charge $-1$ and one electrically neutral particle -- the neutrino. The first generation consists of the Electron, Electron-neutrino, ($\textrm{e}^{-},\nu_{e}$), the second generation consists of the Muon, Muon-neutrino and the ($\mu^{-},\nu_{\mu}$) and the third generation consists of the Tau lepton, Tau-neutrino and the ($\tau^{-},\nu_{\tau}$). These generations are notably  marked by their masses with each member of a higher generation having greater mass than the corresponding particle of the previous generation. Given the above, one wonders at this moment if these equations would explain this mysterious mass hierarchy. This subject will be left for a latter reading.

Equation \ref{cdirac1} possesses symmetric energy solutions, while equations \ref{cdirac2} and \ref{cdirac3} possess asymmetric energy solutions. Given that the asymmetric energy solutions constitute a doublet, one wonders if this doublet could explain the Muon and the Electron, since these particle appear to be similar in nature despite their differences in mass. To answer this question will require that we develop and explore the theory further hence thus, no effort will be made in the present to give an answer to this question. We simple want to point out a possible root to the answer. 

\section{\sectionfont Discussion and Conclusion}

\textbf{First:} The primary aim of this reading has been to show that the anomalous gyromagnetic ratio of the Electron which is not accounted for by the bare Dirac Theory, can be explained as a consequence of the curvature of spacetime. To that end, three Lorentz invariant equations, namely equations \ref{cdirac1}, \ref{cdirac2} and \ref{cdirac3} have been derived. These equations need further investigation regarding their properties such as charge conjugation, time reversal and space inversion. This task has been left for a latter reading.

\textbf{Second:} After discovering the short-comings of the bare Dirac Equation amongst them being how to account for the anomalous gyromagnetic ratio of the Electron, QED had to be developed. In QED, one needs to use Feynman diagrams and make lengthy calculation inorder to explain this ratio. From the view-point of simplicity and beauty, it is difficult to settle for this method as the final program that nature has adopted -- hence the need for a simpler method as has been presented in this expedition. These ideas herein presented, explain this ratio without the need for any Feynman diagram and/or lengthy calculations. 

\textbf{Third:} Besides the simplified explanation of the source of the anomalous gyromagnetic ratio, the metric tensor of spacetime has been reduced to just $4$ vector potentials as compared to the traditional $10$ potentials used in the General Theory o\textit{f} Relativity. To avoid digracing from the main theme of this reading, this $4$ vector potential, has not been defined, as this suffices since we are able to accomplish the primary aim of this reading without actually identifying this field with any of the existing field such as the Electromagnetic field. The task to do so has been left for a latter reading.

\textbf{In closing:} Let me say, from these three equations, that is equations \ref{cdirac1}, \ref{cdirac2} and \ref{cdirac3}, it appears that for the first time, it is possible to find an explanation as to why Leptons exhibit a three stage hierarchy of particles notably  marked by their masses. To keep within the primary aim of the paper, this subject will be left for a latter reading. Certainly more work will have to be done before a clear picture of what these three equations mean thus, this work does not present a complete picture of the Curved Spacetime Dirac Theory. Along the path of furtherwork on the Curved Spacetime Dirac Equation, I could like to say, a second reading is in its final stages of preparation awaiting the publication of the present.
\newpage
%\vspace{2cm}

{\sectionfont \underline{ Acknowledgments}:} \tabletextfont I would like to thank, in alphabetic order of the surname, my friends for their support and encouragement that they have given me in writing this paper: Rita Augustinho, Jotham Dondo, Christina Eddington, Eugene Engelbrecht, Daniel Moeketsi, Anna Neff, Donald Ngobeni and Jasper Snyman. 

%This paper is dedicated to the 280th Anniversary of Sir Isaac Newton's crossing over on 20 March 1727.

%\newpage

\end{document}